# Black Phosphorus/Palladium Nanohybrid: Unraveling the Nature of P-Pd Interaction and Application in Selective Hydrogenation

Matteo Vanni,[a,h] Manuel Serrano-Ruiz,[a] Francesca Telesio,[b] Stefan Heun,[b] Martina Banchelli,[c] Paolo Matteini,[c] Antonio Massimiliano Mio,[d] Giuseppe Nicotra,[d] Corrado Spinella,[d] Stefano Caporali,[e] Andrea Giaccherini,[f] Francesco d'Acapito,[g] Maria Caporali,[a,*] Maurizio Peruzzini[a,*]

[a]CNR-ICCOM, Via Madonna del Piano10, 50019 Sesto Fiorentino, Italy
[b]NEST Istituto Nanoscienze-CNR and Scuola Normale Superiore, Piazza S. Silvestro 12, 56127 Pisa, Italy
[c]CNR-IFAC, Via Madonna del Piano10, 50019 Sesto Fiorentino, Italy
[d]CNR-IMM Istituto per la Microelettronica e Microsistemi, VIII strada 5, I-95121 Catania, Italy.
[e]Department of Industrial Engineering, University of Florence, Via di S. Marta 3, Florence, 50139, Italy
[f]Department of Earth Sciences, University of Florence, Via La Pira 4, Firenze, 50121, Italy
[g]CNR-IOM-OGG, c/o European Synchrotron Radiation Facility - LISA CRG, Grenoble, France.
[h]Department of Biotechnology, Chemistry and Pharmacy, University of Siena, 53100 Siena, Italy

With the advent of molybdenum disulfide and black phosphorus (bP) as emerging 2D materials which have embraced the common interests of physicists, chemists and material scientists, the post-graphene era just started. The peculiar physical properties of black phosphorus are serendipitously in between those of graphene and transition-metal dichalcogenides, as the direct tunable energy band gap (0.3 eV in the bulk, 1.8 eV for the monolayer) and the high carrier mobility at room temperature (1000 $cm^2V^{-1}s^{-1}$), desirable for high-performance mechanically flexible field-effect transistors (FET) devices.[1] bP has a unique puckered honeycomb structure derived from the $sp^3$ hybridization of phosphorus atoms that in turn gives rise to strong in-plane anisotropy of many physical properties such as heat and electron flow which varies according to the armchair and zig-zag direction, respectively.[2] Furthermore, the presence of a lone pair on each P-atom opens the way to several surface modifications. Indeed, several studies have addressed the chemical functionalization of bP, either covalent[3] or not,[4] aiming to improve the processability of the material and the ambient stability,[2-5] which is the Achille's heel of bP. A growing field is the study of bP as support for metal and metal phosphide nanoparticles to trigger applications in catalysis. Taking advantage of its intrinsic property of being a semiconductor, few-layer bP has been tested in photocatalytic processes, either alone or combined with a metal as co-catalyst, in hydrogen evolution reaction (HER),[6] water-splitting[7] and photodegradation of organic pollutants.[8,9] Other catalytic applications were shown for various M NPs/bP nanohybrids where M = Co,[10] Ni,[11] Pt.[12] Recently, Pd NPs were anchored on anatase $TiO_2$-bP hybrid for ethanol electrooxidation.[13] Nevertheless, up to now there is no experimental study on the fundamental interaction between metal nanoparticles and P atoms of bP. Our current work represents the first important step on the elucidation of the nature of this interaction and shows how the interplay between bare palladium nanoparticles and exfoliated black phosphorus makes Pd/bP a highly selective heterogeneous catalyst.

The synthesis of Pd/bP was carried out in solution by direct growth of Pd NPs on exfoliated bP nanosheets obtained by ultrasonication,[14] see ESI. The morphology of the new 2D material was first studied by SEM which shows that Pd NPs are embedded in the bP nanosheets, see Figure 1a. Inspection by TEM, high angle annular dark field scanning transmission electron microscopy (HAADF STEM) and energy dispersive X-ray spectroscopy (EDS) confirmed the formation of nearly spherical palladium nanoparticles homogeneously distributed on the bP nanosheets and with a relative narrow size distribution centered at 3.1±0.8 nm, see Figure 1b-c-d below. Atomic force microscopy (AFM) revealed the presence of both thin and thick flakes with mean size around 600 nm and thickness going from 5 nm to 200 nm (Figure 1e and S4).

**Figure 1.** Structural characterization of Pd/bP. a) SEM image of Pd/bP. b) TEM image of Pd/bP and relative size distribution. Scale

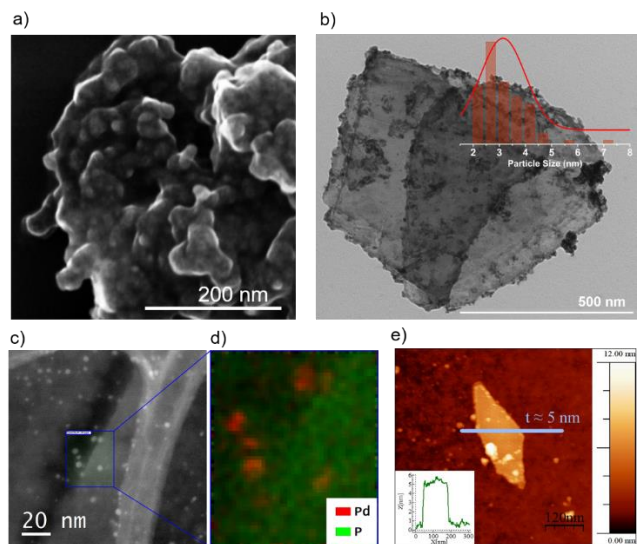

bar: 500 nm. c) High resolution HAADF STEM image of Pd/bP on a lacey carbon grid. Pd-rich areas can be clearly distinguished for the higher Z-contrast (brighter areas). The blue ROI indicates the region in which the EDS SI (spectrum imaging) was performed. (d) EDS elemental mapping of the selected area on Pd/bP obtained integrating the signals from the Pd L-lines and the P K-lines. e) AFM image of a Pd/bP flake on $Si/SiO_2$. The line corresponds to the cross-sectional profile shown as an inset. The flake thickness is approximately 5 nm.

The nanohybrid Pd/bP was characterized by powder X-ray diffraction (PXRD) which confirms the phase purity of the 2D material and shows strong preferential orientation along the (0k0) direction, with three most intense peaks at 2θ = 16.8°, 34.2° and 52.2° assigned respectively to the (020), (040) and (060) planes of orthorhombic black phosphorus, see Figure 2a. This is also characteristic of the pristine material suggesting that its crystalline structure is retained upon functionalization with Pd NPs. A very broad peak around 2θ = 39.5° assigned to the (111) planes of Pd confirms the presence of a nanosized fcc phase of the metal.

It is well established[15] that the three Raman peaks of bP at 357.8, 431.5 and 459.2 cm$^{-1}$, corresponding to the Raman active phonon mode $A^1_g$, $B^2_g$ and $A^2_g$ respectively, are thickness dependent and may undergo a frequency shift with varying flake thickness. Given the thickness polydispersity of bP prepared by sonochemical exfoliation, micro-Raman spectra were collected for a large set of flakes in order to take into account the broad range of thickness. As shown in Figure 2b, Raman spectrum of Pd/bP displays the three peaks characteristic of the orthorombic phase of bP observed above by XRD, however no significant frequency shift was detected compared to pristine bP. Figure S5 shows a detailed statistics for the $A^1_g$, $B^2_g$ and $A^2_g$ Raman modes of pristine bP and Pd/bP.

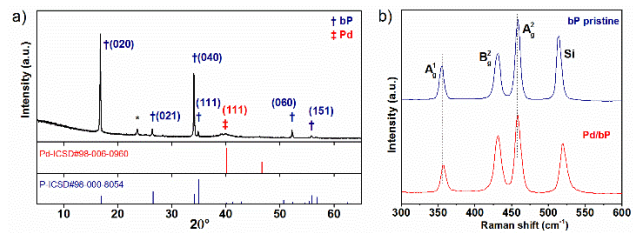

**Figure 2.** a) XRD spectrum of Pd/bP. b) Comparison of Raman spectra collected on pristine bP (red) and Pd/bP (blue). Each spectrum has been obtained combining the data from 15 different flakes.

EELS (Electron Energy Loss Spectroscopy) is an important tool to gain information on the chemical shifts of core-level states, as well as on the fine structure of the unoccupied valence-band states, thus we performed a comparative EELS analysis between bP (simulated[16] and measured) and Pd/bP measured by HAADF-STEM-SI, see Figure 3. An appreciable difference in the $P_L$-edge is observed at around 137 eV for Pd/bP that reveals a modification of the electronic structure of bP which can be interpreted as the result of a strong interaction between P atoms of bP and Pd NPs.

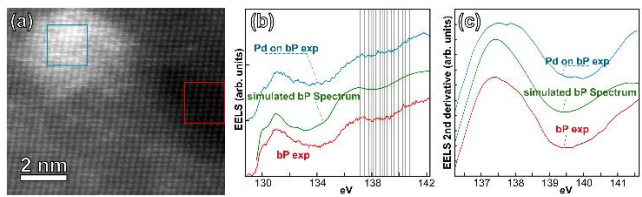

**Figure 3.** (a) HAADF-STEM-SI of Pd/bP acquired along the [101] zone axis, at the P L edge. (b) EELS spectra: red curve, experimental bP, obtained integrating the EELS-SI (spectrum imaging) along the red box in (a); green curve: simulated bP[16]; blue curve: experimental Pd/bP, obtained integrating the EELS-SI along the blue box in (a). (c) second derivative plots of the three EELS spectra within the region highlighted in (b).

The chemical states of the Pd NPs were determined by the Pd 3d core-level XPS spectrum (Figure 4a), which is fitted by the use of two doublets due to spin-orbit couplings. The components at binding energy B.E. = 335.8 eV (Pd $3d_{5/2}$) and at B.E. = 341.0 eV (Pd $3d_{3/2}$) accounts for bulk metallic palladium.[17a] The smaller doublet at B.E. = 338.1 eV and B.E. = 343.5 eV is attributable to an electro-depleted palladium specie, probably with an oxidation state +2. The presence of PdO can be ruled out being absent the peaks with corresponding B.E. values,[17b] thus the signal at higher binding energy can be explained as the result of a partial valence orbital overlap between surface Pd atoms and bP nanosheets. This is consistent with the presence of Pd-P bonds, as deduced from the EELS measurements. The presence of a layer of electron-depleted palladium atoms on the surface of Pd NPs, which strongly interacts with P-atoms, has been already observed[18] in the study of Pd NPs capped with the phosphine ligand PTA (PTA = 1,3,5-triaza-7-phophaadamantane), which indeed show a comparable XPS spectrum.

**Figure 4.** a) Pd 3d core level XPS spectrum of Pd/bP. Metallic and electron-depleted palladium doublets are depicted by blue and red

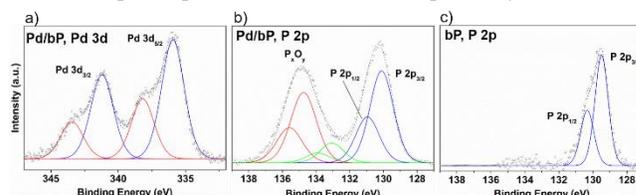

lines, respectively. b) P 2p core level XPS spectrum of Pd/bP, the fitting here requires the use of three different components as discussed in the text. c) P 2p core level XPS spectrum of pristine bP.

The deconvolution of P 2p core level spectrum for pristine bP shows the characteristic doublet representing the P $2p_{3/2}$ and P $2p_{1/2}$ peaks located at 129.7 eV and 130.6 eV, typical of elemental phosphorus, see Figure 4c. Intriguingly, P 2p core level spectrum of Pd/bP results much more complex requiring the use of at least three components. Alongside the doublet attributable to elemental phosphorus (two peaks at B.E.= 130 eV and B.E. = 131 eV), components at higher B.E. values are present as well, see Figure 4b, that can be interpreted as due to phosphorus in different chemical environments. The doublet at intermediate B.E. values (green line fitting in figure 4b) is likely due to P-Pd bonds. The large doublet characterized by the highest binding energy (red line fitting in figure 4b) accounts for the presence of phosphorus oxides species such as P-OH, P=O and P-O-P on the surface of bP sheets.[19] Since peaks at lower BE values characteristic of P having a phosphide nature (128.6 eV)[20] are absent, the presence of phosphides as $Pd_xP_y$ can be excluded.

To the best of our knowledge, the nature of the interaction between metal nanoparticles and bP has been so far approached only by theoretical calculations. For instance, the nanohybrid Ag/bP has been described as stabilized by covalent bonds at the Ag/bP interface by means of DFT calculations[8] but up to now no experimental investigations have been carried out. To unravel the nature of the Pd-P interaction, X-ray absorption spectroscopy (XAS) measurements were performed on Pd/bP and the following compounds were taken as standards: Pd NPs capped with the phosphine ligand PTA and labelled as Pd@PTA,[18] palladium phosphide nanoparticles $PdP_2$ NPs,[21] Pd NPs supported on carbon Pd/C[22] and Pd metal foil, see Table S1.

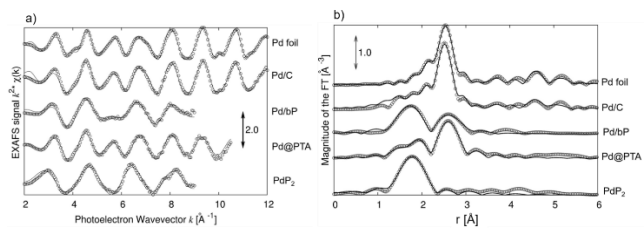

**Figure 5.** a) EXAFS data at the Pd-K edge; b) the corresponding Fourier transforms. Dots are experimental data while continuous lines are the calculated best-fit data.

In Fig. 5, the raw EXAFS spectra and the corresponding Fourier Transforms show clearly the presence of a peak just below 2 Å in the sample Pd/bP that can be attributed to Pd-P bonds having the very similar same shape and almost the same r (Å) value as the peak present in PdP$_2$. A second peak just below 3 Å can be inferred to Pd-Pd bonds, being present both in Pd metal foil, Pd/C and Pd@PTA. This means that Pd atoms are involved in two different bonds: Pd-Pd and Pd-P, the former arising from the bonds with inner metal atoms of the nanoparticle, the latter from an interaction between the surface Pd atoms and P atoms of bP. The precise value of bond lengths can be obtained from these data by fittings and are shown in Table S2. Noteworthy, the short Pd-P distance, $R_{PdP}$ = 2.26(3) Å found in Pd/bP points out that the latter is a very strong interaction. As a comparison, the phosphide PdP$_2$ has longer bonds, $R_{PdP}$ = 2.32(2) Å, as measured by us and according to literature,[23] while PdP$_3$ has[24] $R_{PdP}$ = 2.235 Å which is close to Pd/bP, see Table S2, but the presence of phosphide can be for sure excluded in our sample on the basis of XPS measurements previously discussed. Remarkably, Pd@PTA exhibits $R_{PdP}$ = 2.25 Å, which is consistent with the coordination bond distance Pd-P of 2.203(3) Å measured[25] in the X-ray structure of the complex cation [Pd(PTAH)$_4$]$^{4+}$, and it is almost the same value measured in Pd/bP. This reveals the Pd-P interaction in Pd/bP can be seen as a coordination bond of covalent nature and closely resembles the one exhisting in Pd@PTA: the P atoms of the bP nanosheets surround Pd NPs acting similarly to the molecular phosphine ligand PTA towards Pd NPs. This folding agrees well with the morphology observed by SEM, see Figure1a, and also explains the stacking of the flakes, observed by AFM, see Figure S4. The observed average Pd-P coordination number CNs=1.7(6) in our sample, see ESI for details, fairly corroborates this picture.

To the best of our knowledge, this is the first experimental demonstration that bP nanosheets may act as a polydentate phosphorus ligand towards metal nanoparticles via coordination bonds. This strong interaction, confers a stabilization to Pd NPs preventing their agglomeration and makes Pd/bP a good candidate for catalysis. Intrigued by these results, the chemical process studied was the reduction of chloro-nitroarenes to the corresponding chloroanilines, see Scheme 1. The latter are high valuable intermediates for the manufacture of many agrochemicals, pharmaceuticals, polymers and dyes.[26] Since traditional methods employing stoichiometric reducing agents have drawbacks for both economic and environmental issues, efforts have been devoted to replace them with hydrogen gas. However, many tested catalysts are severely affected by undesired dehalogenation due to the C–Cl hydrogenolysis, (see Figure S6), thus finding a catalyst able to efficiently carry out a chemoselective reduction of nitroarenes remains a challenge.

**Scheme 1.** Hydrogenation of chloro-nitroarenes to chloroanilines.

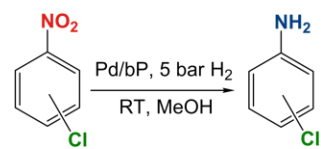

Satisfyingly, Pd/bP showed very high chemoselectivity in the reduction of both *ortho* and *meta*-chloronitrobenzene, reaching 97.3% of selectivity toward the chloroaniline at 99.5% conversion for the *ortho* isomer, see Table S3, overwhelming other heterogeneous catalysts based on Pd NPs[27] that are affected by dehalogenation. As a comparison, Pd NPs with similar average size were grown on a highly porous carbon, ketjen black, see ESI, and at complete conversion the chemoselectivity dropped down to 78.1% as shown in Table S3. Undoubtedly, bP plays a key role since it behaves as a ligand towards Pd NPs, a unique feature that a carbon-based support cannot supply. Additionally, it is relevant that the TOFs are almost the same using as a support either bP or ketjen black, even if the latter is characterized by a very high surface area (1400 m$^2$/g).

Catalyst reuse was also tested, showing the selectivity was maintained unaltered for six consecutive runs, only a small decrease of the conversion was observed, see Figure S7. In agreement with this, TEM investigation on Pd/bP recovered after the catalytic runs, confirmed the morphology was preserved, being the average size of Pd NPs equal to 3.6±0.9 nm, see Figure S8, and the heterogeneous nature of our catalyst was proved as well with an independent test, see Table S4. Thus, the presence of Pd-P bonds prevents bP degradation and provides an excellent structural stability to our catalyst.

In summary, a new Pd/bP nanohybrid has been developed and for the first time, the intimate nature of the interaction between Pd NPs and bP nanosheets was elucidated by EXAFS. A very short Pd-P distance of 2.26(3) Å was disclosed, accounting for a coordinative bond of covalent nature between surface Pd atoms and P atoms. Moreover, the average Pd-P coordination number turned out to be 1.7 (6), which suggests bP acts as a polydentate phosphine ligand towards Pd NPs, stabilizing them toward agglomeration and preventing leaching of the metal in solution. Pd NPs are thus embedded between bP flakes yielding a new 0D-2D heterostructure. The latter was also investigated with other surface techniques, as XPS and EELS-STEM, all of them are consistent with the presence of Pd-P bonds. Finally, the synergy between Pd NPs and bP was successfully exploited in the reduction of chloro-nitroarenes to the corresponding chloro-anilines, showing a far superior chemoselectivity in comparison to other heterogeneous catalysts based on palladium. Furthermore, Pd/bP was reused and maintained its efficiency after six consecutive runs, demonstrating an intrinsic stability that promises further applications in various fields.

## ASSOCIATED CONTENT

**Supporting Information**. The Supporting Information is available free of charge on the ACS Publications website

## AUTHOR INFORMATION

### Corresponding Author

*(M.C.) E-mail: maria.caporali@iccom.cnr.it
ORCID M. Caporali: 0000-0001-6994-7313
*(M.P.) E-mail: maurizio.peruzzini@iccom.cnr.it

ORCID M. Peruzzini: 0000-0002-2708-3964

## Notes
The authors declare no conflict of interest.

## ACKNOWLEDGMENTS
Thanks are expressed to EC for funding the project PHOSFUN "Phosphorene functionalization: a new platform for advanced multifunctional materials" (ERC ADVANCED GRANT to M.P.). STEM-EELS analysis was performed at Beyond-Nano CNR-IMM, which is supported by the Italian Ministry of Education and Research (MIUR) under project Beyond-Nano (PON a3_00363).
The authors gratefully acknowledge for SEM and TEM images "Ce.M.E. – Centro Microscopie Elettroniche Laura Bonzi" in Sesto Fiorentino (Italy) financed by "Ente Cassa di Risparmio di Firenze" and through the projects "EnergyLab", POR FESR 2014-2020 and FELIX (Fotonica ed Elettronica Integrate per l'Industria, project code n. 6455). Finally, MIUR is kindly acknowledged for financial support through Project PRIN 2015 (grant number 20154X9ATP).

# SUPPORTING INFO

# Black Phosphorus/Palladium Nanohybrid: Unraveling the Nature of P-Pd Interaction and Application in Selective Hydrogenation

Matteo Vanni,[a,h] Manuel Serrano-Ruiz,[a] Francesca Telesio,[b] Stefan Heun,[b] Martina Banchelli,[c] Paolo Matteini,[c] Antonio Massilimiliano Mio,[d] Giuseppe Nicotra,[d] Corrado Spinella,[d] Stefano Caporali,[e] Andrea Giaccherini,[f] Francesco d'Acapito,[g] Maria Caporali,*[a] Maurizio Peruzzini*[a]

[a]CNR-Isituto di Chimica dei Composti Organometallici, Via Madonna del Piano10, 50019 Sesto Fiorentino, Italy
[b]NEST Istituto Nanoscienze-CNR and Scuola Normale Superiore, Piazza S. Silvestro 12, 56127 Pisa, Italy
[c]CNR-IFAC, Via Madonna del Piano10, 50019 Sesto Fiorentino, Italy
[d]CNR-IMM Istituto per la Microelettronica e Microsistemi, VIII strada 5, I-95121 Catania, Italy
[e]Department of Industrial Engineering, University of Florence, Via di S. Marta 3, Florence, 50139, Italy
[f]Department of Earth Sciences, University of Florence, Via La Pira 4, Firenze, 50121, Italy
[g]CNR-IOM-OGG, c/o European Synchrotron Radiation Facility - LISA CRG, Grenoble, France
[h]Department of Biotechnology, Chemistry and Pharmacy, University of Siena, 53100 Siena, Italy

## Experimental Section

### 1. General Methods and Materials

All manipulations were performed under inert atmosphere using Schlenk techniques. Tetrahydrofurane (THF) was distilled from sodium/benzophenone and degassed prior to use. $Pd(NO_3)_2 \cdot 2H_2O$, 1-chloro-2-nitrobenzene and 1-chloro-3-nitrobenzene were used as received from Sigma Aldrich. Ketjen black EC 600JD (surface area = 1400 $m^2$/g) was purchased from Akzo Nobel. Black phosphorus (bP) was prepared according to literature procedure[1] and the liquid phase exfoliation of bP was performed according to a protocol set up in our laboratories.[2]

## 2. Synthesis of the nanohybrid Pd/bP

Exfoliated black phosphorus was suspended in degassed THF to have a concentration of 1.0 mg/mL. 1.5 mL of this suspension (1.5 mg of bP, 0.048 mmol) were transferred to a glass vial equipped with a stirring bar and put inside a Schlenk tube. Under stirring, 0.75 mL of degassed ethanol were added, followed by 1.3 mL of an aqueous solution of $Pd(NO_3)_2 \cdot 2H_2O$ (3.75 mM, 0.00487 mmol, molar ratio P : Pd = 10). The vial was transferred in an autoclave, pressurized with 5 bar of $H_2$ and kept on a stirring plate at room temperature for 1 hour. After this time, the autoclave was vented, the catalyst Pd/bP settled down at the bottom of the vial and the colourless supernatant was removed by syringe. 5 mL of degassed ethanol was added and Pd/bP was resuspended by sonication and transferred to a centrifuge tube. After centrifugation at 9000 RPM for 30 minutes the supernatant was discarded and the washing procedure was repeated two times. The solid material was then collected and dried under vacuum for 10 hours. The actual composition of the nanohybrid was evaluated with an inductively coupled plasma atomic emission spectrometer (ICP-AES) measuring both Pd and P. The Pd content resulted 10.6% mol. The yield of this preparation as determined from the final P content was 88%, due to minor material losses during the workup procedure.

The nanohybrid Pd/bP with 0.5 % mol loading of Pd used for comparison in Raman characterization (see Section 4 - Materials Characterization) was prepared following exactly the same procedure.

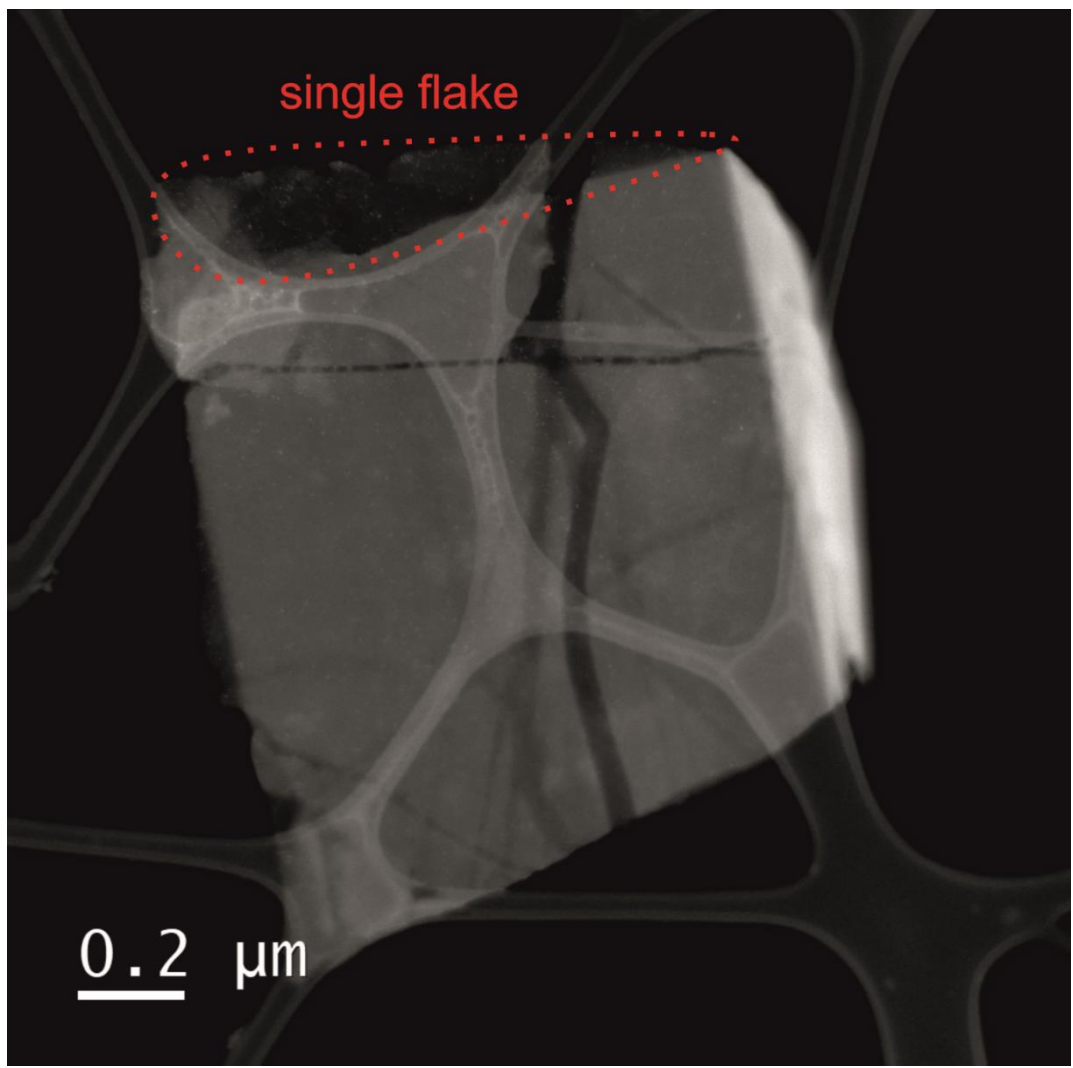

**Figure S1.** Low-magnification HAADF-STEM of Pd/bP on lacey carbon support film. The single flake is clearly visible by Z-contrast.

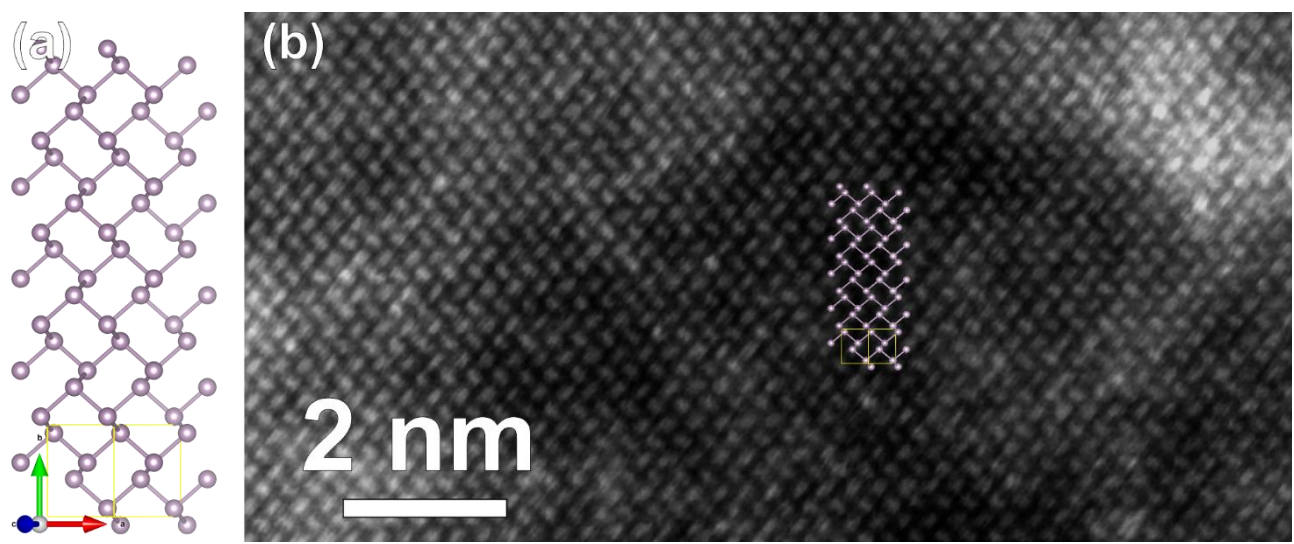

**Figure S2.** (a) Drawing of bP along the [101] zone axis. The yellow frame indicates the unit cell. (b) HAADF-STEM image of Pd/bP taken along the [101] zone axis. The drawing is superimposed to emphasize the correct atomic position. Pd-rich regions can be distinguished for the higher Z-contrast (brighter areas).

## 3. Synthesis of Pd/ketjen black

Ketjen black (0.6 mg, 0.050 mmol) was dispersed in 1.5 mL of distilled and degassed THF by means of ultrasounds. To this black suspension, 1.3 mL of degassed ethanol were added, followed by 1.3 mL of a 3.75 mM aqueous solution of $Pd(NO_3)_2 \cdot 2H_2O$ (0.00487 mmol, Pd : C = 1:10). The mixture was transferred inside a glass vial equipped with a magnetic stirring bar. The vial was put inside an autoclave, pressurized with 5 bar of $H_2$ and left stirring for 1 h. After this time the autoclave was vented, the mixture was transferred in a centrifuge tube, 5.0 mL of degassed ethanol were added and the solid catalyst was isolated upon centrifugation at 9000 rpm for 30 minutes. The washing procedure was repeated two times and the final solid was dried under vacuum for 10 hours. The actual Pd content was measured by ICP-AES and resulted equal to 7.7 % mol.

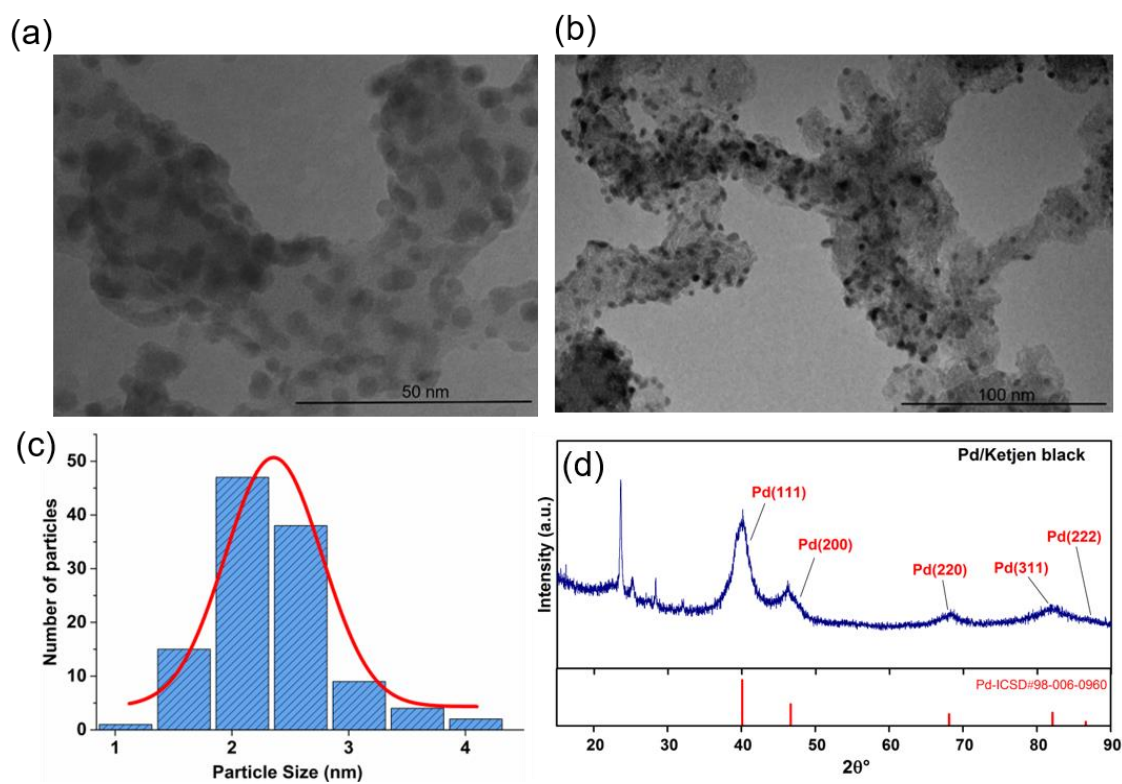

**Figure S3**. Characterization of Pd/ketjen black. (a) Enlarged TEM, scale bar 50 nm and (b) TEM image, scale bar 100 nm (c) Size distribution of PdNPs on ketjen black. (d) XRD pattern of Pd/ketjen black. The peak marked with an asterisk is due to an impurity from the sample holder.

## 4. Materials Characterization

**Transmission electron microscopy.** TEM studies were carried out using a Philips instrument operating at an accelerating voltage of 100 kV. Few drops of Pd/bP and Pd/ketjen black in methanol were placed on the TEM copper/carbon grid, air dried, and measured.

**Scanning transmission electron microscopy.** Atomic-resolution characterization by STEM was performed through a probe aberration-corrected JEOL ARM200CF, equipped with a Ceos hexapole-type Cs corrector, named CESCOR, and operated at a primary beam energy of 60 keV. The electron gun is a cold-field emission gun with an energy spread of 0.3 eV. The probe size was 1.0 Å at 60 kV. Micrographs were acquired in Z-contrast mode (High-Angle Annular Dark Field, HAADF).

A Centurio Energy Dispersive Spectrometer (EDS) equipped with a 100 mm$^2$ Silicon Drift Detector was used for the EDS acquisitions.

A GIF Quantum ER as Electron Energy Loss Spectrometer (EELS) was used for EELS measurements. Both low- and core-loss EELS spectra were acquired with the DualEELS capability through Gatan Digital Micrograph software, which allows the accurate energy calibration of EELS spectra, thanks to the simultaneous alignment of the zero-loss peak position for every single acquisition, which removes any artefact coming from energy shifts. The use of Fourier logarithmic deconvolution on a full spectrum obtained by splicing together low- and core-loss EELS allows removing thickness-related plural scattering.[3] All the STEM-EELS and STEM-EDS measurements were performed simultaneously by using the Gatan spectrum imaging (SI) tool.

**Gas Chromatography.** GC analyses were performed on a Shimadzu GC-14A gas chromatograph (with polar column) equipped with flame ionization detector and a SPB-1 Supelco fused silica capillary column (30 m, 0.25 mm i.d., 0.25 µm film thickness).

**Inductively coupled plasma mass spectrometry.** ICP-MS measurements were performed with an Agilent 7700 Series spectrometer. Samples followed a microwave-assisted digestion in Nitric acid for trace analysis. Then, different dilutions of each sample with water for trace analysis were prepared, in order to obtain concentrations in the sensitivity range of the instrument for the elements under investigation (namely Pd and P). Standards at different concentrations have also been prepared and measured contextually to sample measurements, in order to obtain a calibration curve for each element under investigation.

**Atomic force microscopy.** AFM measurements were performed with a Bruker Dimension Icon Atomic Force Microscope, in pick force mode. Samples for AFM were prepared by drop cast of a suspension of Pd/bP in tetrahydrofuran/ methanol (1:1) on a Si/SiO$_2$ substrate. The drop was left in contact with the substrate for one minute, then the sample was washed with ethanol and dried first under a stream of nitrogen and then in vacuum for five hours. Samples prepared with this method have almost no solvent traces, at least far from the edges, and quite small aggregates. During AFM, we approached regions where no aggregates were visible in the optical microscope, and found some thin and thick structures, the former are shown in the main text, the latter are shown in Figure S4, which are likely to be aggregates of many flakes. These structures have an average lateral dimension of 500 nm and a thickness of up to 200 nm.

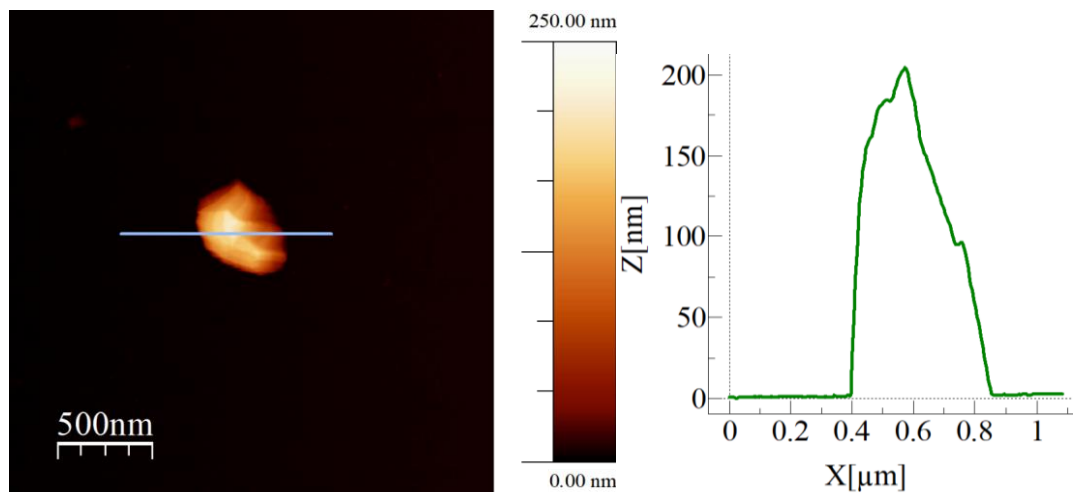

**Figure S4**. AFM image of Pd/bP flake.

**Powder X-ray diffraction.** XRD data were collected with an X'Pert PRO diffractometer operating in a Bragg-Brentano parafocusing geometry with Cu-Kα radiation (λ = 1.5418) operating at 40 kV and 30mA. Samples were prepared by slow dropcast of about 1 mg of material (Pd/bP or Pd/ketjen black) suspended in EtOH (1 mg/mL). A nitrogen flux was directed onto the sample during the dropcast to accelerate evaporation after the deposition of each drop. The process was continued until a layer of the material was uniformly distributed on the sample holder.

**Raman scattering.** Raman measurements were carried out using a micro-Horiba Xplora system coupled to a 532 nm wavelength laser. The backscattered light was collected by a 100× microscope objective with 0.9 NA, which generates a ~1-μm large laser beam waist. Integration times of 10 s, laser power values in the 1-2 mW range and a grating of 1200 cm$^{-1}$ were employed. The samples were prepared by dropcasting a suspension of bP and Pd/bP in tetrahydrofuran on a Si/SiO$_2$ wafer. After one minute of exposure, the wafers were rinsed with ethanol and dried under a stream of nitrogen for 30 minutes. To study the influence of Pd NPs on the Raman shift of black phosphorus, Pd/bP was prepared with two different Pd loading, namely 0.5 % mol and 10% mol, using the same batch of pristine bP for both preparations. Raman spectra were collected from 15 different nanosheets in each sample of pristine bP, Pd/bP 0.5% mol and Pd/bP 10% mol. The average Raman shift of each active mode was calculated for the three samples and is displayed in Figure S5 with relative error bars. No relevant shift in the peak positions was observed for the two Pd/bP samples compared to pristine bP.

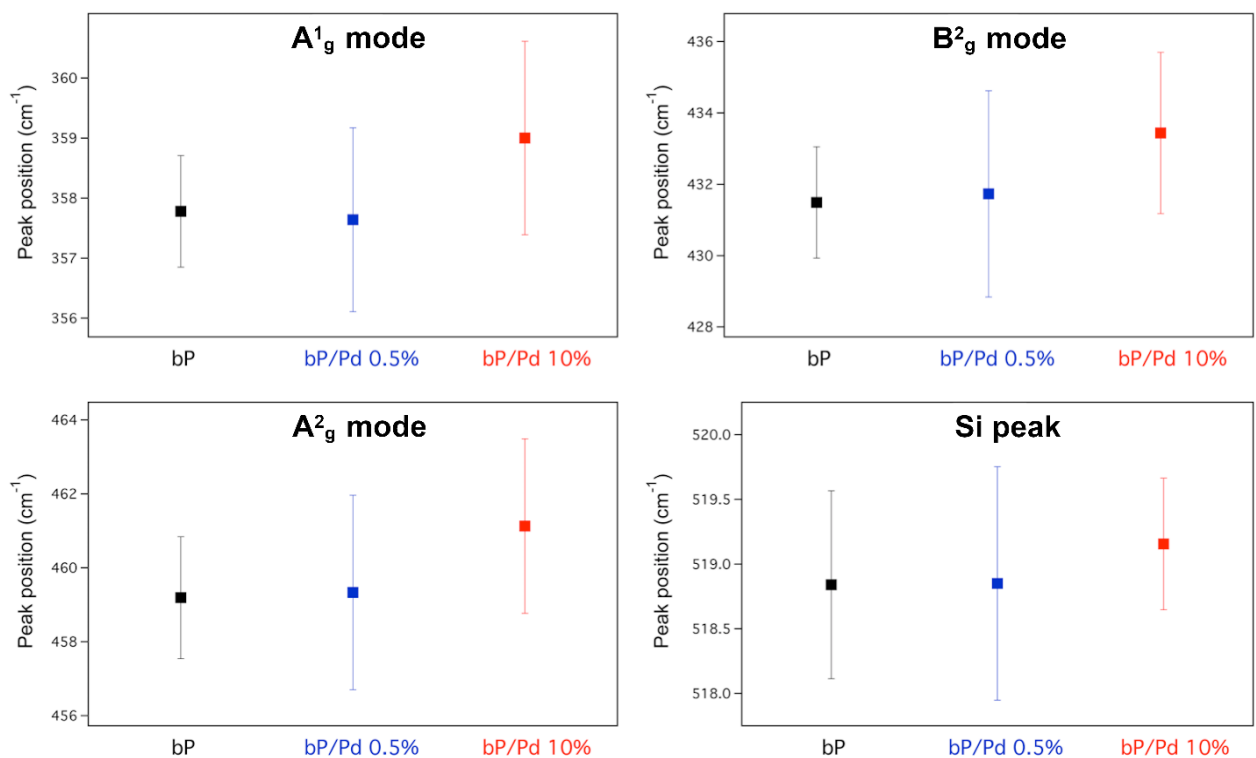

**Figure S5.** Comparison of Raman data of pristine bP with Pd/bP having two different loading of Pd.

**XPS.** X-ray Photoelectron Spectroscopy (XPS) measurements were performed in an ultra-high vacuum ($10^{-9}$ mbar) system equipped with a VSW HAC 5000 hemispherical electron energy analyser and a non-monochromatized Mg-Kα X-ray source (1253.6 eV). The source power used was 100 W (10 kV×10 mA) and the spectra were acquired in the constant-pass-energy mode at $E_{pas}$ = 44 eV. The overall energy resolution was 1.2 eV as a full-width at half maximum (FWHM) for the Ag $3d_{5/2}$ line of a pure silver reference. The recorded spectra were fitted using XPS Peak 4.1 software employing Gauss-Lorentz curves after subtraction of a Shirley-type background. The powder sample was introduced in the UHV system via a loadlock under inert gas ($N_2$) flux, in order to minimize the exposure to air contaminants and kept in the introduction chamber for at least 12 hours before the measurements.

**X-ray Absorption Spectroscopy (XAS).** To study in depth the nature of the interaction between the palladium atoms on the surface of the nanoparticles in contact with P atoms of bP sheets, X-ray Absorption Spectroscopy (XAS) experiments at the Pd-K edge ($E_{edge}$=24350 eV) have been carried out at the LISA beamline at the European Synchrotron Radiation Facility.[4] The monochromator was equipped with a pair of flat Si(311) crystals; collimation and harmonic rejection was achieved by using a Pt-coated cylindrical mirror before the monochromator with an incidence angle of 2 mrad. Further harmonic rejection and focusing was achieved with a Pt toroidal mirror (focusing configuration 2:1) positioned after the monochromator. The beam size on the sample was approximately 0.2 mm. Measurements were carried out at room temperature whereas the XAS signal was collected in fluorescence mode using a 12 elements High Purity Germanium Detector. A Pd foil placed after the sample was used as energy calibration standard compound and its spectrum was always collected together with the samples. XAS data were reduced and analyzed with the ATENA/ARTEMIS codes[5] and the theoretical XAS signals were generated with the FEFF-8.4 code.[6] Structural parameters were obtained by data fits in R space with the transformation ranges in k space varying from case to case (k=[2.5->14] Å$^{-1}$ for for the best cases, k=[2.5->9] Å$^{-1}$ for the noisier spectra) and a k$^2$ weight factor. Coordination Numbers (CNs) have been calibrated via the analysis of the Pd metal foil, finding a global amplitude correction factor $S_0^2$=0.88.

**EXAFS analysis** EXAFS (extended X-ray absorption fine structure) analysis was carried out modeling the data with two contributions: Pd clusters and Pd-P bond. The model of Pd clusters included 4 coordination shells for Pd foil though in Table S2 only the first shell is presented. The others were modelized with a single Pd-Pd first shell, other coordination shells being not visible in the FT. An additional Pd-P bond was added to account for the interaction with phosphorus atoms as evidenced by the peak in the FT below 2 Å. The last three rows of Table S2 show the crystallographic data for different model compounds.

**Table S1**. Description of the samples studied by XAS

| Sample | Description |
| --- | --- |
| Pd/bP | Pd NPs on bP nanosheets (aver. size 3.1±0.8nm) |
| Pd@PTA | PTA-capped Pd NPs (aver. size 3.2 ± 0.4nm) |
| PdP$_2$ | PdP$_2$ NPs |
| Pd/C | Pd NPs on carbon (aver. size 2.0 ± 0.4nm) |
| Pd foil | Pd metal foil at RT |

**Table S2**. Results of the quantitative XAS analysis.

| Sample | CNs Pd-Pd | $R_{PdPd}$(Å) | $\sigma^2_{PdPd}$ (Å$^2$) | CNs Pd-P | $R_{PdP}$(Å) | $\sigma^2_{PdP}$ (Å$^2$) |
|---|---|---|---|---|---|---|
| Pd foil | 12 | 2.74(1) | 0.0059(4) | - | - | - |
| Pd/C | 7(2) | 2.73(1) | 0.0065(5) | 4(2) | 1.94(3) | 0.013(5) |
| Pd/bP | 8(2) | 2.77(3) | 0.016(4) | 1.7(6) | 2.26(3) | 0.0018(6) |
| Pd@PTA[7] | 8(2) | 2.73(2) | 0.009(2) | 0.7(2) | 2.25(3) | 0.004 |
| PdP$_2$ | - | - | - | 3.8(6) | 2.32(2) | 0.004(2) |
| PdP$_2$[8] | - | - | - | 2 | 2.335 | - |
|  |  |  |  | 2 | 2.341 |  |
| PdP$_3$[9] | - | - | - | 6 | 2.235 | - |
| Pd foil[10] | 12 | 2.7453 | - | - | - | - |

CNs $_{PdPd}$ = average coordination number of Pd atoms, $R_{PdPd}$ (Å) = Pd-Pd distance; $\sigma^2_{PdPd}$ (Å$^2$) Debye-Waller factor of the PdPd bond CNs $_{PdP}$ = average coordination number palladium - phosphorus, $\sigma^2_{PdP}$ = Debye-Waller factor of the PdP bond ; $R_{PdP}$ (Å) = Pd-P distance. Errors on the last digit are given in brackets. The numbers in apex, 7, 8, 9, 10 refer to literature references.

## 5. Catalytic hydrogenation of nitroarenes and recycling tests.

In a typical run, 1-chloro-2-nitrobenzene (114.2 mg, 0.725 mmol) was added to the solid catalyst Pd/bP 10% mol (1.3 mg bP, 0.042 mmol, 0.044 mmol Pd) in a screw capped centrifuge tube, 3 mL of degassed MeOH were added and the mixture was sonicated for 5 minutes to suspend the catalyst and dissolve the substrate. The suspension was then transferred in a glass vial placed inside an autoclave and equipped with a magnetic stirring bar. The autoclave was purged with hydrogen (3 times) and then pressurized up to 5 bar. The mixture was kept stirring for the required time, after which the gas was vented and the mixture was transferred in a centrifuge tube. 5 mL of degassed MeOH were added and the suspension was centrifuged at 9000 RPM for 30 minutes. The supernatant was analysed by GC. To the solid residue containing the catalyst, the nitroarene and methanol were added, and a new catalytic run was launched.

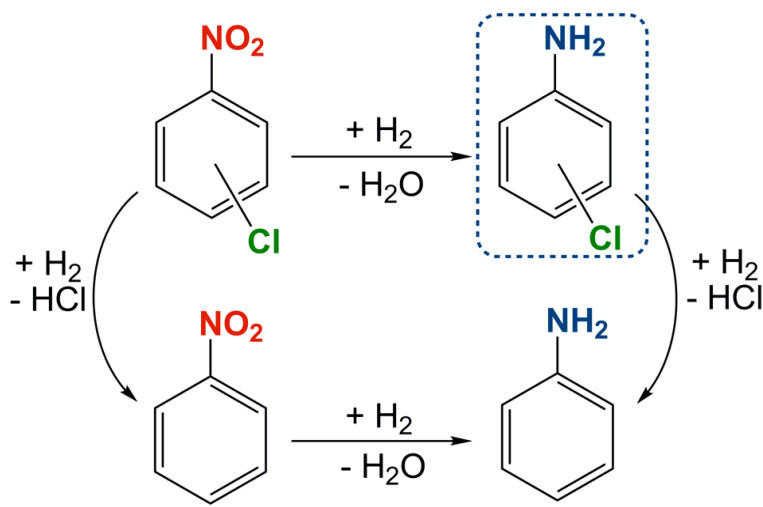

**Figure S6.** The two possible pathways for the secondary reaction of C-Cl hydrogenolysis.

**Table S3.** Catalytic tests on the reduction of nitroarenes.

| Catalyst | Substrate | Time (min) | Sub/cat | Conversion%[a] | Selectivity% | TOF[b] ($h^{-1}$) |
|---|---|---|---|---|---|---|
| Pd/bP | 1-chloro-3-nitrobenzene | 30 | 162 | 99.1 | 97.7 | 313 |
|  | 1-chloro-2-nitrobenzene | 40 | 162 | 99.5 | 97.3 | 235 |
| Pd/C | 1-chloro-2-nitrobenzene | 30 | 191 | 99.9 | 78.1 | 298 |

Reaction conditions: solvent: methanol, [substrates] = 0.242 M; room temperature; 5 bar $H_2$; magnetic stirring, 1200 rpm. Sub/Cat: substrate/catalyst ratio (mol/mol). [a]Data from GC analysis and GC-MS; [b] Turn over frequency

## 6. Test for the heterogeneous nature of the catalyst, ruling out the presence of catalytically active species in solution.

A catalytic run was started under the standard reaction conditions described above. After 15 minutes the reaction was stopped, the autoclave was vented and the reaction mixture was transferred in a centrifuge tube. The solid material was isolated by centrifugation at 9000 RPM for 30 minutes. A small fraction of the supernatant (2 µL) was taken for GC-MS analysis, Entry 1 in Table S3. The remaining was transferred in the autoclave, pressurized with 5 bar of $H_2$ and kept stirring for 40 minutes. After this time the autoclave was vented and GC-MS analysis was carried out. The supernatant gave no further conversion (Entry 2), confirming that the catalytic activity is due entirely to the solid phase.

A further catalytic run was performed recycling the recovered solid catalyst under the same reaction conditions. Results comparable with fresh catalyst Pd/bP were obtained, see Entry 3.

**Table S4**.

| Entry | Catalytic run | Time (min) | Conversion% | Selectivity% | TOF (h$^{-1}$) |
|---|---|---|---|---|---|
| 1 | 1$^{st}$ run, Pd/bP | 15 | 42.6 | 95.1 | 263 |
| 2 | no solid phase | 40 | 42.8 | 94.7 | - |
| 3 | 1$^{st}$ recycle, Pd/bP | 40 | 97.8 | 90.7 | 215 |

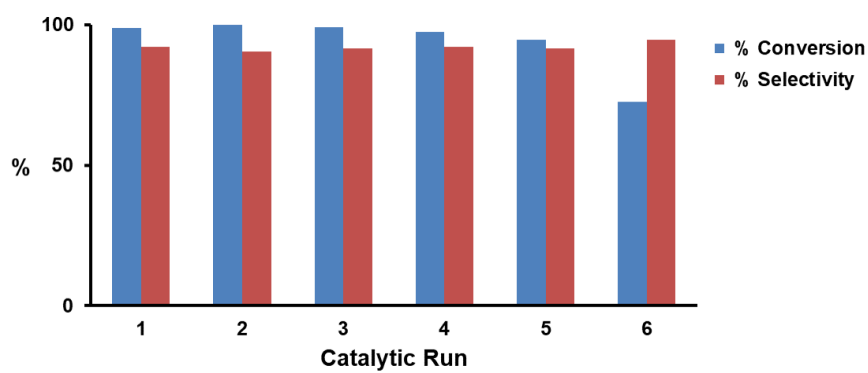

**Figure S7**. Catalyst reuse in the hydrogenation of 1-chloro-2-nitrobenzene.

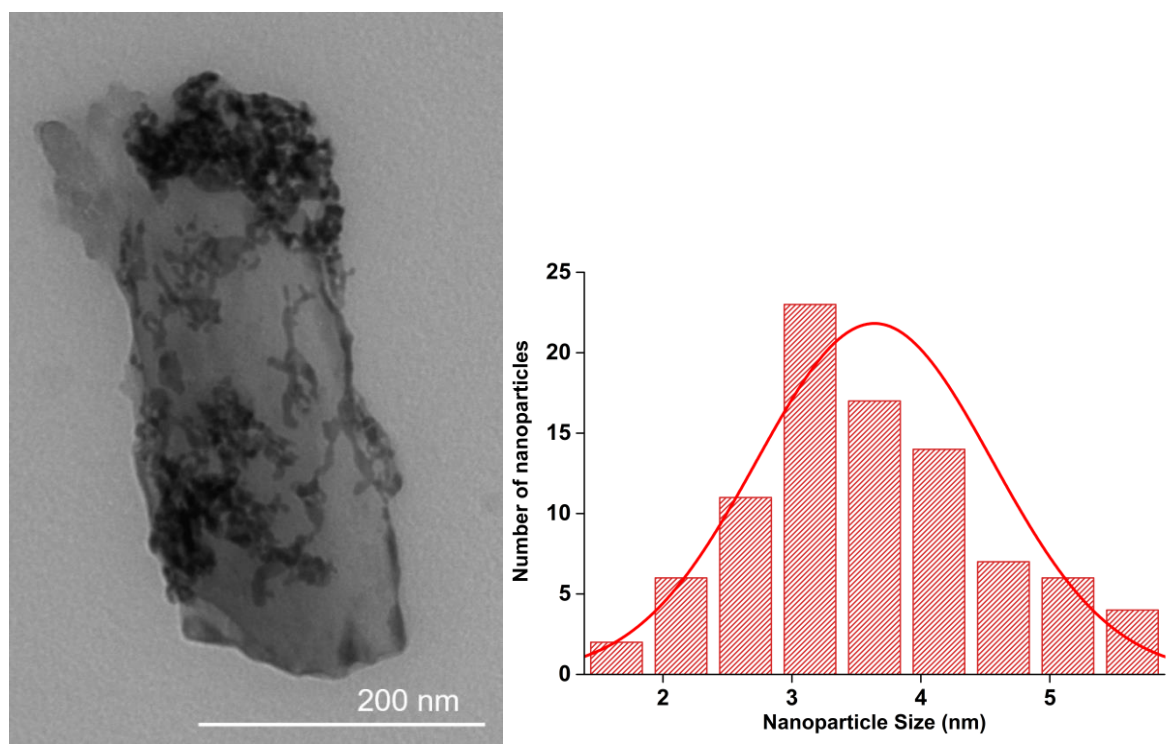

**Figure S8**. TEM image of Pd/bP after catalytic tests and relative size distribution of the palladium nanoparticles.